\documentclass[12pt]{article}
\newfont{\bbd}{msbm10 scaled\magstep1}
\def\C{\hbox{\bbd C}}
\def\R{\hbox{\bbd R}}
\def\beq{\begin{equation}} \def\eeq{\end{equation}}
\def\be{\begin{displaymath}} \def\ee{\end{displaymath}}
\def\bea{\begin{eqnarray}} \def\eea{\end{eqnarray}}
\def\beas{\begin{eqnarray*}} \def\eeas{\end{eqnarray*}}
 \def\e{{\rm e}} \def\phi{\varphi}

 \def\tr{\mathop{\hbox{\rm tr}}\nolimits}

\newcommand{\tfrac}[2]{{\textstyle\frac{#1}{#2}}}
\renewcommand{\theequation}{\thesection.\arabic{equation}}
\newcounter{subequation}[equation]
\makeatletter

\expandafter\let\expandafter
\reset@font\csname reset@font\endcsname

\def\subeqnarray{\arraycolsep1pt
    \def\@eqnnum\stepcounter##1{\stepcounter{subequation}%
        {\reset@font\rm(\theequation\alph{subequation})}}
\jot5mm     \eqnarray}

\makeatother

\begin{document}

\renewcommand{\thefootnote}{\fnsymbol{footnote}}
\begin{center} {\Large \bf  Inverse problem for sl(2) lattices}
\end{center}
\renewcommand{\thefootnote}{\arabic{footnote}}
\vskip0.2cm
\renewcommand{\thefootnote}{\fnsymbol{footnote}}
\begin{center} {\bf Vadim B.~Kuznetsov\footnote[2]{EPSRC
Advanced Research Fellow}}
\end{center}
\renewcommand{\thefootnote}{\arabic{footnote}}
\vskip 0.2cm
\begin{center} Department of Applied Mathematics\\ University of Leeds,
LEEDS LS2 9JT, United Kingdom\\
E-mail: V.B.Kuznetsov@leeds.ac.uk
\end{center}
\vskip0.5cm
\begin{center}
{\bf Abstract} \end{center}
\vskip0.5cm
\noindent
We consider the inverse problem for periodic sl(2) lattices
as a canonical transformation from the separation to local
variables. A new concept of a {\it factorized separation chain}
is introduced allowing to solve the inverse problem explicitly.
The method is applied to an arbitrary representation of the
corresponding Sklyanin algebra.
\vskip 6cm
\pagebreak

\section{Introduction}
\setcounter{equation}{0}
Let us denote a separating transform
from the local variables $({\bf q},{\bf p})$ to the separation
variables $({\bf u},{\bf v})$
as ${\mathcal S}_n$,
\beq
{\mathcal S}_n:\qquad ({\bf q},{\bf p})\mapsto ({\bf u},{\bf v}),
\eeq
and the inverse separating map as  ${\mathcal S}_n^{-1}$,
\beq
{\mathcal S}_n^{-1}:\qquad ({\bf u},{\bf v})\mapsto ({\bf q},{\bf p}).
\eeq

In \cite{S1,S2} and \cite{S3} Sklyanin
worked out the $r$-matrix technique for
the method of separation of variables.
It was based on the 1976 separation of
the (classical) periodic Toda lattice by
van Moerbeke \cite{M} and  Flaschka--McLaughlin
\cite{FM} and also on early 80's Gutzwiller's solution to
the quantum periodic Toda lattice \cite{Gutz}
and Komarov's ideas \cite{Kom} on the relation between the quantum
separation of variables and the quantum inverse scattering
method. In those three papers, Sklyanin studied separating maps ${\mathcal S}_n$
for the Goryachev-Chaplygin top, periodic Toda lattice and
Heisenberg magnet, respectively. He also
studied the corresponding quantum separations $\hat{\mathcal S}_n$.

Further development of the separation method included:
(i) the 3-particle elliptic Calogero-Moser system \cite{Skl}
and its trigonometric and $q$- versions \cite{KS95,KS96,KNS}
for which {\it both} maps, $\hat{\mathcal S}_3$ and $\hat{\mathcal S}_3^{-1}$,
were explicitly constructed, and (ii) recent construction
of the direct map $\hat{\mathcal S}_4$ in the case of the
4-particle Calogero-Moser system \cite{Man}.

In his Nankai Lectures \cite{S3} Sklyanin announced a
new method (Lecture 4) that allowed to reduce
the $n$-variable spectral problem for the kernel
of the inverse separating map $\hat{\mathcal S}_n^{-1}$
to (smaller) $p$- and $(n-p)$-variable spectral
problems.

Recently, Kharchev and Lebedev \cite{Lebed} used this
method to describe the quantum separating map
$\hat{\mathcal S}_n^{-1}$ for the Toda lattice.

In the present paper we show that, in the classical
case, Sklyanin's approach
is equivalent to explicit solution of the inverse problem
for sl(2) integrable lattices. We introduce a
new concept of the (inverse) {\it factorized separation chain}:
\beq
{\mathcal S}_n^{-1}={\mathcal B}_{1}\circ\cdots\circ{\mathcal B}_{n-1}\circ{\mathcal B}_n.
\label{chain}
\eeq
Each of the factors ${\mathcal B}_m$
is constructed as a map between the separation variables of the
$m$-particle lattice and those of the $(m-1)$-particle lattice, plus
a pair of initial (local) variables $(q_m,p_m)$. By definition,
the map ${\mathcal B}_m$ is the same as the map ${\mathcal B}_n$
with $n$ replaced by $m$, so that it acts only on $m$ degrees
of freedom. This means that the whole chain (\ref{chain})
is generated by a recursive application of a single map
${\mathcal B}_n$.

Such factorization of the inverse
separating map is explicitly performed for an arbitrary
representation of the quadratic Sklyanin algebra in the
considered case of the $2\times 2$ Lax matrix.
It represents an interesting
algebraic structure of the intertwiner between the separation
and lattice representations of the quadratic algebra.
It also reveals a hidden algebraic structure of the
inverse problem for integrable lattices.

In fact, the factorization (\ref{chain})
is a direct consequence of the multiplicative structure of the
 Lax (monodromy) matrix.
Using this structure of the Lax matrix,
we show that the corresponding  transformation ${\mathcal B}_n$:
(i) exists, (ii) is unique and (iii) is a rational canonical
map with a simple generating function.
We derive such
factorization first for the generic case of the inhomogeneous
Heisenberg magnet. The formulae for two degenerate cases,
the periodic DST and Toda lattices, are also given.

The structure of the paper is following. In Section 2
we define our basic model. Sections 3 and 4 give the
separation and lattice representations of the
quadratic algebra, respectively. Section 5 describes
the recursive procedure for
solving the inverse problem. In Section 6
we find the generating function of the factorized
separation map ${\mathcal B}_n$.
In Sections 7 and 8 we apply our results to the DST
and periodic Toda lattice.
There are three Appendices
with explicit formulae for the inverse separating maps
in the case of 3 spins/particles.

%
%
\section{Quadratic $r$-matrix algebra and the model}
\setcounter{equation}{0}
We study a class of finite-dimensional Liouville
integrable systems described by the
representations of the quadratic $r$-matrix Poisson algebra,
or the Sklyanin algebra:
\beq
\{\stackrel{1}{L}(u),\stackrel{2}{L}(v)\}=r(u-v)\,\stackrel{1}{L}(u)\stackrel{2}{L}(v)\,
-\stackrel{1}{L}(u)\stackrel{2}{L}(v)\,\, r(u-v).
\label{1}\eeq
We consider the simplest case of the $4\times 4$  rational $r$-matrix $r(u)$
and $2\times2$ $L$-operator $L(u)$:
\beq
r(u)=\frac{\kappa}{u}\,\pmatrix{1&0&0&0\cr 0&0&1&0\cr 0&1&0&0\cr 0&0&0&1},
\qquad L(u)=\pmatrix{A(u)&B(u)\cr C(u)&D(u)}.
\eeq
In (\ref{1}) the standard notations for tensor products are used:
\begin{equation}\label{2}
\stackrel{1}{L}(u)=L(u)\otimes \pmatrix{1&0\cr 0&1},\quad
\stackrel{2}{L}(v)=\pmatrix{1&0\cr 0&1}\otimes L(v).
\end{equation}
In the left-hand side of (\ref{1}) one has a $4\times 4$
matrix with the Poisson brackets as its entries:
$(\{\stackrel{1}{L}(u),\stackrel{2}{L}(v)\})_{ij,kl}$ $\equiv\{(L(u))_{ij},(L(v))_{kl}\}$.
In the right-hand side of (\ref{1}) there are ($4\times4$) matrix
products.

Sklyanin's bracket (\ref{1}) amounts to having the following Poisson brackets
between the elements $A(u)$, $B(u)$, $C(u)$ and $D(u)$ of the
$L$-operator $L(u)$:
\begin{equation}\label{3}
\{A(u),A(v)\}=\{B(u),B(v)\}=\{C(u),C(v)\}=\{D(u),D(v)\}=0,
\end{equation}
\begin{equation}\label{4}
\{B(u),A(v)\}=\frac{\kappa}{u-v}\left(B(u)A(v)-B(v)A(u)\right),
\end{equation}
\begin{equation}\label{5}
\{C(u),A(v)\}=\frac{\kappa}{u-v}\left(A(u)C(v)-A(v)C(u)\right),
\end{equation}
\begin{equation}\label{6}
\{B(u),D(v)\}=\frac{\kappa}{u-v}\left(D(u)B(v)-D(v)B(u)\right),
\end{equation}
\begin{equation}\label{7}
\{C(u),D(v)\}=\frac{\kappa}{u-v}\left(C(u)D(v)-C(v)D(u)\right),
\end{equation}
\begin{equation}\label{8}
\{A(u),D(v)\}=\frac{\kappa}{u-v}\left(C(u)B(v)-C(v)B(u)\right),
\end{equation}
\begin{equation}\label{9}
\{B(u),C(v)\}=\frac{\kappa}{u-v}\left(D(u)A(v)-D(v)A(u)\right).
\end{equation}

We choose the entries of our $L$-operator $L_n(u)$ to be polynomials
of degree $n$:
\bea
L_n(u)&=&\pmatrix{A_n(u)&B_n(u)\cr C_n(u)&D_n(u)}\label{14-10}\\
&=&\pmatrix{
\alpha u^{n}+A_{n,1}u^{n-1}+\ldots+A_{n,n}&
\beta u^{n}+B_{n,1}u^{n-1}+\ldots+B_{n,n}\cr
\gamma u^{n}+C_{n,1}u^{n-1}+\ldots+C_{n,n}&
\delta u^{n}+D_{n,1}u^{n-1}+\ldots+D_{n,n}}.
\nonumber\eea
Notice that the leading coefficients, $\alpha$, $\beta$, $\gamma$, $\delta$,
are Casimirs of the bracket (\ref{1}). By another property of the bracket,
there are $2n$ further Casimirs $Q_k$, $k=1,\ldots,2n$,
which are coefficients of the $\det L_n(u)$:
\beq
\det L_n(u)=(\alpha\delta-\beta\gamma)u^{2n}+Q_{1}u^{2n-1}+\ldots+Q_{2n}.
\eeq
Therefore, we have a $4n$-dimensional space of the coefficients
$A_{n,i}, B_{n,i},C_{n,i},D_{n,i}$ of
the matrix $L_n(u)$ with $2n$ Casimir operators, leaving us with $n$
degrees of freedom. Independent, Poisson involutive integrals of motion
$H_i$, $i=1,\ldots,n$, are given by the coefficients of the $\tr L_n(u)$:
\beq
\tr L_n(u)=(\alpha+\delta)u^{n}+H_{1}u^{n-1}+\ldots+H_n,\quad
\{H_i,H_j\}=0.
\eeq
These define a Liouville integrable system which is our generic model
for the whole paper. The two first Hamiltonians of the system are
\beq\label{hams}
H_{1}=A_{n,1}+D_{n,1},\qquad H_2=A_{n,2}+D_{n,2}.
\eeq
%
%
\section{Separation representation}
\setcounter{equation}{0}
Our first aim is to construct a separation representation for the
quadratic algebra (\ref{3})--(\ref{14-10}). In this special representation
one has $n$ canonical pairs of variables, $u_i$, $v_i$, $i=1,\ldots,n$,
having the standard Poisson brackets,
\beq
\{u_i,u_j\}=\{v_i,v_j\}=0,\qquad \{v_i,u_j\}=\delta_{ij},
\label{12}
\eeq
with the $u$-variables being $n$ zeros of the polynomial $B_n(u)$
and the $\e^{\kappa v}$-variables being values of the polynomial $A_n(u)$
at those zeros,
\beq
B_n(u_i)=0,\qquad \e^{\kappa v_i}=A_n(u_i),\qquad i=1,\ldots,n,
\label{11}\eeq
so that the pairs $(\e^{\kappa v_i},u_i)$, $i=1,\ldots,n$, belong to the spectral
curve ${\mathcal C}_n$ of the $L$-operator $L_n(u)$:
\beq
{\mathcal C}_n=\{(\e^{\kappa v},u)\in\C^2|\det(L_n(u)-\e^{\kappa v})=0\},
\;\; \det(L_n(u_i)-\e^{\kappa v_i})=0.
\eeq

Let us parameterize the determinant of the $L$-operator as follows:
\beq
\det L_n(u)=(\alpha\delta-\beta\gamma)\prod_{i=1}^n
\left((u-c_i)^2+\kappa^2s_i^2\right).
\eeq
The interpolation data (\ref{11}) plus $n+1$ identities, $A_n(u_i)D_n(u_i)=\det L_n(u_i)$
and $\det L_n(u)$ $=A_n(u)D_n(u)-B_n(u)C_n(u)$, allow us to construct the needed
separation representation for the whole algebra:
\bea
B_n(u)&=&\beta(u-u_1)(u-u_2)\cdots(u-u_n),\label{19}\\
A_n(u)&=&B_n(u)\left(\frac{\alpha}{\beta}+\sum_{i=1}^n \frac{\e^{\kappa v_i}}{(u-u_i)B_n^{'}(u_i)}\right),\\
D_n(u)&=&B_n(u)\left(\frac{\delta}{\beta}+\sum_{i=1}^n \frac{\det L_n(u_i)\;\e^{-\kappa v_i}}
{(u-u_i)B_n^{'}(u_i)}\right),\\
C_n(u)&=&\frac{A_n(u)D_n(u)-\det L_n(u)}{B_n(u)}\,.
\label{20}
\eea
One can easily check that the brackets (\ref{3})--(\ref{9}) imply the brackets (\ref{12})
and vice versa.
%
%
\section{Lattice representation}
\setcounter{equation}{0}
Another important representation of the quadratic algebra with the generators
$A_{n,i},B_{n,i},$ $C_{n,i}$ and $D_{n,i}$ comes as a consequence of the
co-multiplication property of the algebra (\ref{1}). Essentially, it means that
the $L$-operator (\ref{14-10}) can be factorized into
a product of elementary matrices, each containing only one degree of freedom.
In this picture, our main model turns out to be an $n$-site Heisenberg
magnet, which is an integrable lattice of $n$ sl(2) spins with nearest
neighbour interaction.

In this lattice representation the $L$-operator (\ref{14-10}) acquires
the following form:
\beq\label{factors}
L_n(u)=\ell_n(u-c_n)\;\ell_{n-1}(u-c_{n-1})\;\cdots\;\ell_1(u-c_1)\;\pmatrix{\alpha&\beta\cr
\gamma&\delta},
\eeq
\beq
\ell_i(u):=\pmatrix{u-\mbox{\rm i}\kappa s_i^{(3)}& -\mbox{\rm i}\kappa s_i^{(-)}\cr
 -\mbox{\rm i}\kappa s_i^{(+)}&
u+\mbox{\rm i}\kappa s_i^{(3)}},\qquad i=1,\ldots,n.
\eeq
The local variables ${\bf s}_i$, $i=1,\ldots,n$, are generators of $n$ copies
of the sl(2) Poisson algebra:
\beq
\{s^{(3)},s^{(\pm)}\}=\mp\mbox{\rm i}s^{(\pm)},\qquad
\{s^{(-)},s^{(+)}\}=2\mbox{\rm i}s^{(3)}.
\eeq
The (non-local) Hamiltonians $H_1$ and $H_2$ (\ref{hams})
are as follows:
\bea
H_1&=&-\mbox{\rm i}\kappa\sum_{i=1}^n\left((\alpha-\delta)s_i^{(3)}
+\beta s_i^{(+)}+\gamma s_i^{(-)}\right)-(\alpha+\delta)\sum_{i=1}^nc_i,\\
H_2&=&-\kappa^2\sum_{i>j}\left[\alpha\left(s_i^{(3)}s_j^{(3)}+s_i^{(-)}s_j^{(+)}\right)
+\beta\left(s_i^{(+)}s_j^{(3)}-s_i^{(3)}s_j^{(+)}\right)
 \right.\\
&&\left.\qquad\qquad +\;\gamma\left(s_i^{(3)}s_j^{(-)}-s_i^{(-)}s_j^{(3)}\right)
+\delta\left(s_i^{(3)}s_j^{(3)}+s_i^{(+)}s_j^{(-)}\right)\right]\nonumber\\
&& +\;\sum_{j\neq i}c_j\left(\tfrac{\alpha+\delta}{2}\,c_i+\mbox{\rm i}
\kappa(\alpha-\delta)s_i^{(3)}+\mbox{\rm i}\kappa\beta s_i^{(+)}
+\mbox{\rm i}\kappa \gamma s_i^{(-)}\right).
\nonumber\eea

We use the following `holomorphic' representation
for the local spins ${\bf s}_i$, and correspondingly for the local
$L$-operators $\ell_i(u)$, in terms of $n$ pairs of canonical
variables $\bf q$ and $\bf p$:
\beq
\ell_i(u)=\pmatrix{u+\mbox{\rm i}\kappa\left(\mbox{\rm i}q_ip_i-s_i\right)
& -\mbox{\rm i}\kappa p_i\cr
 -\mbox{\rm i}\kappa\left(q_i^2p_i+2\mbox{\rm i}s_iq_i\right)&
u-\mbox{\rm i}\kappa\left(\mbox{\rm i}q_ip_i-s_i\right)},\quad i=1,\ldots,n,
\eeq
\beq
\{q_i,q_j\}=\{p_i,p_j\}=0,\qquad \{p_i,q_j\}=\delta_{ij}.
\label{15}
\eeq
Here, the variables $s_i$ are the spin values: ${\bf s}_i^2
=\left(s_i^{(3)}\right)^2+s_i^{(+)}s_i^{(-)}=s_i^2$, $i=1,\ldots,n$.
One can easily check that the brackets (\ref{3})--(\ref{9}) imply the brackets (\ref{15})
and vice versa.

Our aim is to describe in explicit terms the transformation
between the two representations outlined in this and in the
previous Section, notably in the direction from the
variables $\bf u$ and $\bf v$ to the local spins ${\bf s}_1,\ldots,{\bf s}_n$
or to the variables  $\bf q$ and $\bf p$.
In other words, we present below a constructive solution
to the inverse problem which provides the formulae for
the local variables in terms of separation variables for a large class
of integrable lattices.
%
%
\section{Inverse problem}
\setcounter{equation}{0}
The formulae (\ref{11}) are the defining equations for the symplectic
separating map ${\mathcal S}_n$
from the initial (local) variables to the separation variables. A natural
question arises whether this map can be inverted in a sensible way.
The answer is affirmative.

The map ${\mathcal S}_n$ and its inverse are rather complicated
canonical transforms, meaning that it is difficult to write down their
corresponding generating functions
or to quantize these maps starting directly from the defining
equations (\ref{11}). Our prime motivation therefore is
to solve the inverse problem explicitly in such a way that consequent
quantization is straightforward.

In order to construct our solution to the inverse problem we shall
introduce a new concept of a {\it factorized separation chain}.
It means that the inverse separating
transform ${\mathcal S}_n^{-1}$ will be factorized into a composition (a chain) of
elementary canonical transforms\footnote{an analogous factorization chain
exists for the direct separating map ${\mathcal S}_n$ but this is beyond the scope of this paper}.
The meaning of the factors can be understood from considering the first map
in the chain, ${\mathcal B}_n$, which, in a sense, represents a transformation
between the model with $n$ degrees of freedom and the model with $n-1$
degrees of freedom. Consequent iteration of the found map will then result in
the needed inverse separating transform ${\mathcal S}_n^{-1}$:
\beq
{\mathcal S}_n^{-1}={\mathcal B}_{1}\circ\cdots\circ{\mathcal B}_{n-1}\circ{\mathcal B}_n.
\eeq

Let us construct a canonical transformation that factorizes out
the $n$th local $L$-operator $\ell_n(u-c_n)$ from the $n$-spin
$L$-operator $L_n(u)$ (cf. (\ref{factors})). This is the transformation ${\mathcal B}_n$.
That is to say that we have the following matrix equation:
\bea
L_n(u)
&=&\pmatrix{u-c_n+\mbox{\rm i}\kappa\left(\mbox{\rm i}q_np_n-s_n\right)
& -\mbox{\rm i}\kappa p_n\cr
 -\mbox{\rm i}\kappa\left(q_n^2p_n+2\mbox{\rm i}s_nq_n\right)&
u-c_n-\mbox{\rm i}\kappa\left(\mbox{\rm i}q_np_n-s_n\right)}\times\quad\label{30}\\
&&\times\; L_{n-1}(u),
\nonumber\eea
where the $n$-spin matrix on the left is given in terms of separation
variables $u_i$, $v_i$, $i=1,\ldots,n$, by interpolation
formulae (\ref{19})--(\ref{20}) and the $(n-1)$-spin matrix on the right
is given by similar formulae in terms of its own separation
variables\footnote{N.B.: $n-1$ canonical pairs/degrees of freedom}
$\tilde u_j$, $\tilde v_j$, $j=1,\ldots,n-1$:
\bea
B_{n-1}(u)&=&\beta(u-\tilde u_1)(u-\tilde u_2)\cdots(u-\tilde u_{n-1}),\label{21}\\
A_{n-1}(u)&=&
  B_{n-1}(u)\left(\frac{\alpha}{\beta}
+\sum_{j=1}^{n-1} \frac{\e^{\kappa \tilde v_j}}{(u-\tilde u_j)B_{n-1}^{'}(\tilde u_j)}\right),\\
D_{n-1}(u)&=&B_{n-1}(u)\left(\frac{\delta}{\beta}+
\sum_{j=1}^{n-1} \frac{\det L_{n-1}(\tilde u_j)\;\e^{-\kappa \tilde v_j}}
{(u-\tilde u_j)B_{n-1}^{'}(\tilde u_j)}\right),\\
C_{n-1}(u)&=&\frac{A_{n-1}(u)D_{n-1}(u)-\det L_{n-1}(u)}{B_{n-1}(u)}\,.
\label{22}
\eea

Transformation ${\mathcal B}_n$ maps separation variables $u_i$, $v_i$, $i=1,\ldots,n$,
into new separation variables $\tilde u_j$, $\tilde v_j$, $j=1,\ldots,n-1$, which
parameterize the matrix $L_{n-1}(u)$, and into a pair of local variables, $q_n$ and $p_n$.
It is a single-valued rational map. Indeed, rewrite the matrix equation (\ref{30}) in the
equivalent form,
\beq
L_{n-1}(u)=\frac{1}{\det \ell_n(u-c_n)} \;\ell_n^\wedge(u-c_n) L_n(u),
\label{31}\eeq
where $\ell_n^\wedge(u-c_n)$ stands for the adjoint matrix.
Equating to zero the residues of the right-hand side at two zeros
$c_n\pm\mbox{\rm i}\kappa s_n$ of the $\det \ell_n(u-c_n)$
and solving the resulting equations in $q_n$ and $p_n$, we obtain
\beq
q_n=-\mbox{\rm i}\,\frac{D_n(c_n-\mbox{\rm i}\kappa s_n)}{B_n(c_n-\mbox{\rm i}\kappa s_n)}\,,
\eeq
\beq
p_n=\frac{2s_nB_n(c_n+\mbox{\rm i}\kappa s_n)B_n(c_n-\mbox{\rm i}\kappa s_n)}
{B_n(c_n+\mbox{\rm i}\kappa s_n)D_n(c_n-\mbox{\rm i}\kappa s_n)
-B_n(c_n-\mbox{\rm i}\kappa s_n)D_n(c_n+\mbox{\rm i}\kappa s_n)}\,.
\eeq
This gives the answer for the $n$th pair of local variables.
Now, the formula (\ref{31}) uniquely defines the matrix $L_{n-1}(u)$
in terms of the separation variables $\mbox{\bf u}$ and $\mbox{\bf v}$.
The procedure can be recursively repeated,
leading to reconstruction of all local variables $\bf q$ and $\bf p$. For instance,
explicit formulae for the inverse separating map ${\mathcal S}_3^{-1}$
for the homogeneous chain of three $0$-spins,
i.e. when all parameters vanish, $s_i=c_i=0$,
are given in the Appendix A.
The corresponding formulae for the case of non-zero parameters
$s_i$ and $c_i$ are too long.
%
%
\section{Generating function}\setcounter{equation}{0}
In the previous Section we introduced the factorized separation chain
${\mathcal S}_n^{-1}={\mathcal B}_{1}\circ\cdots\circ{\mathcal B}_{n-1}\circ{\mathcal B}_n$
that maps the canonical variables in the following order:
\beq\label{mm}
{\mathcal S}_n^{-1}:
({\bf u},{\bf v})\stackrel{{\mathcal B}_n}{\mapsto}(\tilde {\bf u}, \tilde {\bf v}|q_n,p_n)%
\stackrel{{\mathcal B}_{n-1}}{\mapsto}(\tilde{\tilde {\bf u}},\tilde{\tilde {\bf v}}|q_{n-1},p_{n-1};q_n,p_n)%
\cdots\stackrel{{\mathcal B}_1}{\mapsto} ({\bf q}, {\bf p}).
\eeq
In this inverse separating chain, the initial variables $({\bf u}, {\bf v})$ are
the separation variables for the starting $n$-spin integrable
system and the terminal variables $({\bf q}, {\bf p})$ are the local
variables. The intermediate variables $(\tilde {\bf u}, \tilde {\bf v})$ have the meaning
of being the separation variables for the $(n-1)$-spin system, the
variables $(\tilde{\tilde {\bf u}}, \tilde{\tilde {\bf v}})$ being those
for the $(n-2)$-spin system, and so on. Notice that every next
transformation in the chain acts only on the new separation pairs from
the outcome of a previous transform, without touching
the pairs $(q_i,p_i)$ of local variables that have already been factorized out,
so that the transform
${\mathcal B}_m$ acts on less and less variables (degrees of freedom)
as its index $m$ decreases.

In this Section we show that the factors ${\mathcal B}_m$, $m=1,\ldots,n$,
 of the composition
(\ref{mm}) are very simple canonical transforms and we describe them in completely
explicit terms by presenting their generating functions. The complexity
of the compound transform ${\mathcal S}_n^{-1}$ is therefore being explained
as {\it a composition} of these elementary transforms.  Such factorization
of a complex separating transform must be a universal feature of all
separating maps
and it must provide a unified approach to their description.
The application of this approach is not limited by the model
chosen in this paper.

Let us start from the first map ${\mathcal B}_n$ and let us fix the $2n$ variables of
its generating function $F_n$ to be the coordinates $\bf u$ and, respectively,
$\tilde{\bf u}$ and $q_n$, so that one has the following $2n$ equations defining the map
${\mathcal B}_n$,
\bea
{\mathcal B}_n:&&\quad v_i=\frac{\partial F_n(\tilde{ {\bf  u}},q_n|{\bf u})}{\partial u_i}\,,\qquad
i=1,\ldots,n,\\
&&\quad \tilde v_j=-\frac{\partial F_n(\tilde{ {\bf  u}},q_n|{\bf u})}{\partial \tilde u_j}\,,\quad
j=1,\ldots,n-1,\\
&&\quad p_n=-\frac{\partial F_n(\tilde{ {\bf  u}},q_n|{\bf u})}{\partial  q_n}\,.
\eea
We will first find these equations and then find the function $F_n(\tilde{ {\bf  u}},q_n|{\bf u})$.

Equation for the polynomial $B_n(u)$ from the equality (\ref{30}) reads
\beq\label{hihi3}
B_n(u)=(u-c_n-\mbox{\rm i}\kappa s_n-\kappa q_np_n)B_{n-1}(u)
-\mbox{\rm i}\kappa p_n D_{n-1}(u).
\eeq
Substituting $u=\tilde u_j$, one obtains
\beq\label{hihi1}
B_n(\tilde u_j)=-\mbox{\rm i}\kappa p_n D_{n-1}(\tilde u_j),\qquad j=1,\ldots,n-1.
\eeq

Equation for the polynomial $B_{n-1}(u)$ from the equality (\ref{31}) reads
\beq
\det\left(\ell_n(u-c_n)\right) B_{n-1}(u)
=(u-c_n+\mbox{\rm i}\kappa s_n+\kappa q_np_n)B_{n}(u)
+\mbox{\rm i}\kappa p_n D_{n}(u).
\eeq
Substituting $u=u_i$, one obtains
\beq\label{hihi2}
\det\left(\ell_n(u_i-c_n)\right) B_{n-1}(u_i)
=\mbox{\rm i}\kappa p_n D_{n}(u_i),\qquad i=1,\ldots,n.
\eeq

Equating the leading coefficients in $u$ in (\ref{hihi3}) and unwrapping
the formulae (\ref{hihi1}) and (\ref{hihi2}), one finally obtains
$2n$ needed equations for the map ${\mathcal B}_n$ in the
following form:
\bea\label{bn1}
p_n&=&\frac{\beta}{\kappa}\,\frac{\sum_{i=1}^nu_i-\sum_{j=1}^{n-1}\tilde u_j
-c_n-\mbox{\rm i}\kappa s_n}{\beta q_n+\mbox{\rm i}\delta}\,,\\
\e^{\kappa v_i}&=&\frac{\mbox{\rm i}\kappa p_n(\alpha\delta-\beta\gamma)}
{\beta}\prod_{m=1}^{n-1}\frac{(u_i-c_m)^2+\kappa^2s_m^2}{u_i-\tilde u_m}\,,
\quad i=1,\ldots,n,\label{bn2}\\
\e^{\kappa \tilde v_j}&=&\frac{-\mbox{\rm i}\kappa p_n(\alpha\delta-\beta\gamma)}
{\beta(\tilde u_j-u_n)}\prod_{m=1}^{n-1}\frac{(\tilde u_j-c_m)^2+\kappa^2s_m^2}
{\tilde u_j-u_m}\,, j=1,\ldots,n-1.\quad\label{bn3}
\eea
These equations can be easily integrated, resulting in explicit
formula for the generating function $F_n(\tilde{ {\bf  u}},q_n|{\bf u})$
of the map ${\mathcal B}_n$:
\bea\label{gen-fun}
F_n(\tilde{ {\bf  u}},q_n|{\bf u})&=&
\frac{1}{\kappa}\,\sum_{i=1}^n\sum_{j=1}^{n-1}\Omega\left(\tilde u_j-u_i\right)\\
&&-\frac{1}{\kappa}\,\Omega\left(\sum_{j=1}^{n-1}\tilde u_j-\sum_{i=1}^nu_i
+c_n+\mbox{\rm i}\kappa s_n\right)\nonumber\\
&&-\frac{1}{\kappa}\,\left(\sum_{j=1}^{n-1}\tilde u_j-\sum_{i=1}^nu_i\right)
\log\frac{\mbox{\rm i}(\alpha\delta-\beta\gamma)}{\beta q_n+\mbox{\rm i}\delta}\nonumber\\
&&+\frac{1}{\kappa}\,\left(c_n+\mbox{\rm i}\kappa s_n\right)
\log\left(\beta q_n+\mbox{\rm i}\delta\right)-\frac{\mbox{\rm i}\pi n}{\kappa}\,\sum_{i=1}^nu_i
\nonumber\\
&&+\frac{1}{\kappa}\,\sum_{i=1}^n\sum_{j=1}^{n-1}
\left(\Omega\left(u_i-c_j+\mbox{\rm i}\kappa s_j\right)
     +\Omega\left(u_i-c_j-\mbox{\rm i}\kappa s_j\right)\right)\nonumber\\
&&-\frac{1}{\kappa}\,\sum_{j=1}^{n-1}\sum_{j'=1}^{n-1}
\left(\Omega\left(\tilde u_j-c_{j'}+\mbox{\rm i}\kappa s_{j'}\right)
     +\Omega\left(\tilde u_j-c_{j'}-\mbox{\rm i}\kappa s_{j'}\right)\right).
\nonumber\eea
Here, the function $\Omega(u)$ is the anti-derivative
of $\log(u)$:
\beq
\Omega(u)=\int^u\log(u') \, du'=u\left(\log(u)-1\right)+C,\qquad \frac{d\Omega(u)}{du}=\log(u).
\eeq

In order to obtain the corresponding formulae for the maps ${\mathcal B}_m$, $m=1,\ldots,n-1$,
one must replace $n$ by $m$ in the formulae (\ref{bn1}), (\ref{bn2}), (\ref{bn3}) and
(\ref{gen-fun}). It was already mentioned that a reduction in the number
of variables happens along with decreasing of map's index.

The formulae in this Section completely describe the inverse separating map
${\mathcal S}_n^{-1}$ as a canonical map, through the explicit representation for its
factors.
In the next two Sections we consider two degenerate cases of the
Heisenberg magnet: the integrable DST model and the periodic
Toda lattice. The main reason for inserting these Sections
is to exemplify the new concept
of the factorized separation chain that has been introduced above.
For these two degenerate cases we give definitions of the models,
equations and generating functions
for the basic map ${\mathcal B}_n$ and, finally,
analogues of the rational map ${\mathcal S}_3^{-1}$
in the 3-particle case.
%
%
\section{Integrable DST model}\setcounter{equation}{0}
The integrable case of the DST (discrete self-trapping) model
with $n$ degrees of freedom was introduced in \cite{CJK} and
studied in \cite{KSS}. It appears as a specialization of our
basic model when several parameters vanish:
\beq
\beta=\gamma=\delta=0\qquad \mbox{\rm and}\qquad Q_j=0, \qquad j=1,\ldots,n-1.
\eeq
We also put $\alpha=1$, leading to the $L$-operator
\bea
L_n(u)&=&\pmatrix{
u^{n}+A_{n,1}u^{n-1}+\ldots+A_{n,n}&
B_{n,1}u^{n-1}+\ldots+B_{n,n}\cr
C_{n,1}u^{n-1}+\ldots+C_{n,n}&
D_{n,2}u^{n-2}+\ldots+D_{n,n}}.\quad
\label{dst14-10}\eea
Notice that $D_{n,1}\equiv Q_1=0$. Set $Q_n=b^n$
and parameterize $\det L_n(u)$ as follows:
\beq
\det L_n(u)=b^n(u-c_1)(u-c_2)\cdots (u-c_n).
\eeq
Let us also choose $\kappa=-1$. By definition, the $L$-operator (\ref{dst14-10})
obeys the quadratic relations of the Poisson algebra (\ref{3})--(\ref{9}).

Separation variables $(\e^{-v_i},u_i)$ are introduced as before:
\beq
B_n(u_i)=0,\qquad \e^{-v_i}=A_n(u_i),\qquad i=1,\ldots,n-1,
\label{dst11}\eeq
the only difference now is that this gives only $n-1$
instead of $n$ separation pairs. The missing pair of
canonical variables is defined as follows:
\beq\label{dst12}
v_n:=B_{n,1},\qquad u_n:=\frac{A_{n,1}}{B_{n,1}}\,.
\eeq
We remind that the $n$ pairs introduced by (\ref{dst11}) and
(\ref{dst12}) are canonical variables, i.e.
\beq
\{u_i,u_j\}=\{v_i,v_j\}=0,\qquad \{v_i,u_j\}=\delta_{ij},\qquad
i,j=1,\ldots,n.
\label{dst13}
\eeq
The separation representation of the algebra in this special case
has the form
\bea
B_n(u)&=&v_n(u-u_1)(u-u_2)\cdots(u-u_{n-1}),\label{dst19}\\
A_n(u)&=&B_n(u)\left(\frac{u+u_nv_n+\sum_{i=1}^{n-1}u_i}{v_n}+\sum_{i=1}^{n-1}
\frac{\e^{-v_i}}{(u-u_i)B_n^{'}(u_i)}\right),\\
D_n(u)&=&B_n(u)\,\sum_{i=1}^{n-1} \frac{\det L_n(u_i)\;\e^{v_i}}
{(u-u_i)B_n^{'}(u_i)}\,,\\
C_n(u)&=&\frac{A_n(u)D_n(u)-\det L_n(u)}{B_n(u)}\,.
\label{dst20}
\eea

In the lattice representation, the $L$-operator (\ref{dst14-10}) acquires
the form
\beq\label{dstfactors}
L_n(u)=\ell_n(u-c_n)\;\ell_{n-1}(u-c_{n-1})\;\cdots\;\ell_1(u-c_1),
\eeq
with the local $L$-operators
\beq
\ell_i(u):=\pmatrix{u-q_ip_i& bq_i\cr -p_i&b},\qquad i=1,\ldots,n.
\eeq
The local variables $(q_i,p_i)$ are canonical,
\beq
\{q_i,q_j\}=\{p_i,p_j\}=0,\qquad \{p_i,q_j\}=\delta_{ij},\qquad
i,j=1,\ldots,n.
\label{dst15}
\eeq
The Hamiltonians $H_1$ and $H_2$ from the
$\tr L_n(u)=u^{n}+H_{1}u^{n-1}+\ldots+H_n$ are
\bea
H_1&=&-\sum_{i=1}^n(q_ip_i+c_i),\\
H_2&=&\sum_{i>j}(q_ip_i+c_i)(q_jp_j+c_j)
-b\sum_{i=1}^nq_{i+1}p_i,\qquad\quad
(q_{n+1}\equiv q_1).
\nonumber\eea

The (inverse) factorized separation chain
${\mathcal S}_n^{-1}={\mathcal B}_{1}\circ\cdots\circ{\mathcal B}_{n-1}\circ{\mathcal B}_n$
is produced by a recursive application of the map ${\mathcal B}_n$,
which is defined by the following equations:
\bea\label{dstbn1}
v_n&=&\tilde v_{n-1}=\tfrac{\sum_{j=1}^{n-2}\tilde u_j-\sum_{i=1}^{n-1}u_i}
{u_n-\tilde u_{n-1}}\,,\quad p_n=\tfrac{\sum_{i=1}^{n-1}u_i-\sum_{j=1}^{n-2}\tilde u_j
-c_n}{q_n}\,,\qquad\\
\e^{- v_i}&=&-\frac{b^nq_n(u_i-c_{n-1})}
{v_n}\prod_{m=1}^{n-2}\frac{u_i-c_m}{u_i-\tilde u_m}\,,
\qquad i=1,\ldots,n-1,\label{dstbn2}\\
\e^{- \tilde v_j}&=&\frac{b^nq_n}{v_n}
\prod_{m=1}^{n-1}\frac{\tilde u_j-c_m}
{\tilde u_j-u_m}\,,
\qquad\qquad\qquad\quad\; j=1,\ldots,n-2.\label{dstbn3}
\eea
These equations define a canonical map
with the generating function
\bea\label{dstgen-fun}
F_n(\tilde{ {\bf  u}},q_n|{\bf u})&=&
\sum_{i=1}^{n-1}\sum_{j=1}^{n-2}\Omega\left(u_i-\tilde u_j\right)
+\Omega\left(\sum_{i=1}^{n-1}u_i-\sum_{j=1}^{n-2}\tilde u_j\right)\\
&&-\left(\sum_{i=1}^{n-1}u_i-\sum_{j=1}^{n-2}\tilde u_j\right)
\log\left(b^n(u_n-\tilde u_{n-1})\right)-\mbox{\rm i}\pi n\sum_{j=1}^{n-2}\tilde u_j\nonumber\\
&&-\sum_{i=1}^{n-1}\sum_{i'=1}^{n-1}\Omega\left(u_i-c_{i'}\right)
+\sum_{i=1}^{n-1}\sum_{j=1}^{n-2}\Omega\left(\tilde u_j-c_{i}\right)\nonumber\\
&&+\left(\sum_{j=1}^{n-2}\tilde u_j-\sum_{i=1}^{n-1}u_i+c_n\right)\log q_n.
\nonumber\eea
Explicit expressions for the $n$th pair of local variables are
\beq
q_n=\frac{B_n(c_n)}{D_n(c_n)}\,,\qquad p_n=-\frac{D_{n,2}}{v_n}=-\sum_{i=1}^{n-1}
 \frac{\det L_n(u_i)\;\e^{v_i}}{B_n^{'}(u_i)}\,.
\eeq
This process can be iterated, leading to explicit rational formulae for the
inverse separating map ${\mathcal S}_n^{-1}$. The corresponding formulae
in the homogeneous ($c_i=0$) 3-particle
case are given in the Appendix B.
%
%
\section{Periodic Toda lattice}\setcounter{equation}{0}
The periodic Toda lattice appears as a further specialization of our
basic model when the parameters are fixed as follows:
\beq
\beta=\gamma=\delta=0\qquad \mbox{\rm and}\qquad Q_k=0, \qquad k=1,\ldots,2n-1.
\eeq
We also put $\alpha=1$ (and $\kappa=-1$), leading to the $L$-operator which is
similar to the one for the DST lattice:
\bea
L_n(u)&=&\pmatrix{
u^{n}+A_{n,1}u^{n-1}+\ldots+A_{n,n}&
B_{n,1}u^{n-1}+\ldots+B_{n,n}\cr
C_{n,1}u^{n-1}+\ldots+C_{n,n}&
D_{n,2}u^{n-2}+\ldots+D_{n,n}}.\quad
\label{tldst14-10}\eea
Set $Q_{2n}=1$, so that
\beq
\det L_n(u)=1.
\eeq

Separation variables $(\e^{-v_i},u_i)$, $i=1,\ldots,n$, are introduced by the same
formulae as for the DST
lattice, cf. (\ref{dst11}) and (\ref{dst12}).
The separation representation of the quadratic algebra in this case
reads
\bea
B_n(u)&=&v_n(u-u_1)(u-u_2)\cdots(u-u_{n-1}),\label{tldst19}\\
A_n(u)&=&B_n(u)\left(\frac{u+u_nv_n+\sum_{i=1}^{n-1}u_i}{v_n}+\sum_{i=1}^{n-1}
\frac{\e^{-v_i}}{(u-u_i)B_n^{'}(u_i)}\right),\\
D_n(u)&=&B_n(u)\,\sum_{i=1}^{n-1} \frac{\e^{v_i}}
{(u-u_i)B_n^{'}(u_i)}\,,\\
C_n(u)&=&\frac{A_n(u)D_n(u)-1}{B_n(u)}\,.
\label{tldst20}
\eea

In the lattice representation, the $L$-operator (\ref{tldst14-10}) acquires
the form
\beq\label{tldstfactors}
L_n(u)=\ell_n(u)\;\ell_{n-1}(u)\;\cdots\;\ell_1(u)
=\pmatrix{A_n(u)&B_n(u)\cr C_n(u)&D_n(u)},
\eeq
with the local $L$-operators
\beq
\ell_i(u):=\pmatrix{u-p_i& \e^{q_i}\cr -\e^{-q_i}&0},\qquad i=1,\ldots,n.
\eeq
The Hamiltonians $H_1$ and $H_2$ from the
$\tr L_n(u)=u^{n}+H_{1}u^{n-1}+\ldots+H_n$ are
\beq
H_1=-\sum_{i=1}^np_i,\quad
H_2=\sum_{i>j}p_ip_j
-\sum_{i=1}^n\e^{q_{i+1}-q_i},\quad
(q_{n+1}\equiv q_1).
\eeq

The (inverse) factorized separation chain
${\mathcal S}_n^{-1}={\mathcal B}_{1}\circ\cdots\circ{\mathcal B}_{n-1}\circ{\mathcal B}_n$
is produced by a recursive application of the map ${\mathcal B}_n$,
which is defined by the following equations:
\bea\label{tldstbn1}
v_n&=&\tilde v_{n-1}=\frac{\sum_{j=1}^{n-2}\tilde u_j-\sum_{i=1}^{n-1}u_i}
{u_n-\tilde u_{n-1}}\,,\quad p_n=\sum_{i=1}^{n-1}u_i-\sum_{j=1}^{n-2}\tilde u_j,
\quad\\
\e^{v_i}&=&-v_n\e^{-q_n}
\prod_{m=1}^{n-2}\left(u_i-\tilde u_m\right),
\qquad\quad i=1,\ldots,n-1,\label{tldstbn2}\\
\e^{\tilde v_j}&=&v_n\e^{-q_n}
\prod_{m=1}^{n-1}\left(\tilde u_j-u_m\right),
\qquad\quad\;\; j=1,\ldots,n-2.\label{tldstbn3}
\eea
These equations define a canonical map
with the generating function
\bea\label{tldstgen-fun}
F_n(\tilde{ {\bf  u}},q_n|{\bf u})&=&
\sum_{i=1}^{n-1}\sum_{j=1}^{n-2}\Omega\left(u_i-\tilde u_j\right)
+\Omega\left(\sum_{i=1}^{n-1}u_i-\sum_{j=1}^{n-2}\tilde u_j\right)\\
&&-\left(\sum_{i=1}^{n-1}u_i-\sum_{j=1}^{n-2}\tilde u_j\right)
\left(q_n+\log (u_n-\tilde u_{n-1})\right)
-\mbox{\rm i}\pi n\sum_{j=1}^{n-2}\tilde u_j.
\nonumber\eea

Explicit expressions for the $n$th pair of local variables are
\beq
\e^{-q_n}=-\tfrac{D_{n,2}}{v_n}=-\sum_{i=1}^{n-1}
 \tfrac{\e^{v_i}}{B_n^{'}(u_i)}\,,\;
p_n=\sum_{i=1}^{n-1}u_i+\tfrac{D_{n,3}}{D_{n,2}}
=\tfrac{\sum_{i=1}^{n-1}\tfrac{u_i\e^{v_i}}{B_n^{'}(u_i)}}
{\sum_{i=1}^{n-1}\tfrac{\e^{v_i}}{B_n^{'}(u_i)}}\,.
\eeq
This process can be iterated. The corresponding formulae
for the 3-particle periodic Toda lattice are given in the Appendix C.
%
%
\section*{Acknowledgments}
A part of this work was done in November 2000 when
the author visited D\'epartement de Math\'ematiques, Universit\'e
de Poitiers, France. I want to thank Pol Vanhaecke for hospitality and for
valuable discussions. The results were presented
at the SPT Workshop, May 19--26, 2002, in Cala Gonone, Sardinia.
The support of the EPSRC is gratefully acknowledged.
%
%
\section*{Appendix A}
\renewcommand{\theequation}{A.\arabic{equation}}
\setcounter{equation}{0}
Here we give explicit formulae for the inverse map ${\mathcal S}_3^{-1}$
for the chain of three spins when $s_i=c_i=0$, $i=1,2,3$.
For notation, see the end of Section 5. Denote $w_i=\e^{\kappa v_i}$,
$i=1,2,3$, then
\bea
p_1&=&\frac{-\mbox{\rm i}\beta}{(\alpha\delta-\beta\gamma)\kappa}\,
\tfrac{{\left|\matrix{w_1&w_2&w_3\cr u_1&u_2&u_3\cr u_1^2&u_2^2&u_3^2}\right|}^2}
{  \left|\matrix{1&1&1\cr u_1&u_2&u_3\cr u_1^2&u_2^2&u_3^2}\right| \,
 \left|\matrix{w_1&w_2&w_3\cr u_1^2&u_2^2&u_3^2\cr u_1^3&u_2^3&u_3^3}\right|     }\,,
\label{p1}\\
p_2&=&\frac{-\mbox{\rm i}\beta}{(\alpha\delta-\beta\gamma)\kappa}\,
\tfrac{{\left|\matrix{u_1w_1&u_2w_2&u_3w_3\cr w_1&w_2&w_3\cr u_1^3&u_2^3&u_3^3}\right|}^2}
{  \left|\matrix{w_1&w_2&w_3\cr u_1^2&u_2^2&u_3^2\cr u_1^3&u_2^3&u_3^3}\right| \,
 \left|\matrix{u_1w_1&u_2w_2&u_3w_3\cr w_1&w_2&w_3\cr u_1^4&u_2^4&u_3^4}\right|     }\,,
\label{p2}\\
p_3&=&\frac{\mbox{\rm i}\beta w_1w_2w_3}{(\alpha\delta-\beta\gamma)\kappa}\,
\tfrac{\left|\matrix{1&1&1\cr u_1&u_2&u_3\cr u_1^2&u_2^2&u_3^2}\right|}
{  \left|\matrix{u_1w_1&u_2w_2&u_3w_3\cr w_1&w_2&w_3\cr u_1^4&u_2^4&u_3^4}\right|}\,,
\label{p3}\\
q_1&=&-\mbox{\rm i}\,\frac{\delta}{\beta}+
\frac{\mbox{\rm i}(\alpha\delta-\beta\gamma)u_1u_2u_3}{\beta}\,
\tfrac{\left|\matrix{1&1&1\cr u_1&u_2&u_3\cr u_1^2&u_2^2&u_3^2}\right|}
{  \left|\matrix{w_1&w_2&w_3\cr u_1&u_2&u_3\cr u_1^2&u_2^2&u_3^2}\right|}\,,
\label{q1}\\
q_2&=&-\mbox{\rm i}\,\frac{\delta}{\beta}+
\frac{\mbox{\rm i}(\alpha\delta-\beta\gamma)}{\beta}\,
\tfrac{\left|\matrix{w_1&w_2&w_3\cr u_1^3&u_2^3&u_3^3\cr u_1^4&u_2^4&u_3^4}\right|}
{  \left|\matrix{u_1w_1&u_2w_2&u_3w_3\cr w_1&w_2&w_3\cr u_1^3&u_2^3&u_3^3}\right|}\,,
\label{q2}\\
q_3&=&-\mbox{\rm i}\,\frac{\delta}{\beta}-
\frac{\mbox{\rm i}(\alpha\delta-\beta\gamma)}{\beta w_1w_2w_3}\,
\tfrac{\left|\matrix{u_1w_1&u_2w_2&u_3w_3\cr w_1&w_2&w_3\cr u_1^5&u_2^5&u_3^5}\right|}
{  \left|\matrix{1&1&1\cr u_1&u_2&u_3\cr u_1^2&u_2^2&u_3^2}\right|}\,.
\label{q3}\eea
%
%
\section*{Appendix B}
\renewcommand{\theequation}{B.\arabic{equation}}
\setcounter{equation}{0}
Here we give explicit formulae for the inverse map ${\mathcal S}_3^{-1}$
for the 3-particle DST lattice when $c_i=0$, $i=1,2,3$ (cf. Section 7).
Denote $w_i=\e^{-v_i}$, $i=1,2$, then
\bea
p_1&=&-bu_3-\frac{b}{u_1u_2v_3}\,
\tfrac{\left|\matrix{w_1-u_1^3&w_2-u_2^3\cr u_1&u_2}\right|}
{  \left|\matrix{1&1\cr u_1&u_2}\right|     }\,,\qquad\;\;
q_1=\frac{v_3}{b}\,,\\
p_2&=&\frac{b^2}{v_3}+\frac{b^2u_1^2u_2^2}{v_3}\,
\tfrac{\left|\matrix{1&1\cr u_1&u_2}\right|}
{  \left|\matrix{w_1&w_2\cr u_1^2&u_2^2}\right|}\,,
\qquad
q_2=\frac{v_3}{b^2u_1u_2}\,
\tfrac{\left|\matrix{w_1&w_2\cr u_1&u_2}\right|}
{  \left|\matrix{1&1\cr u_1&u_2}\right|}\,,
\\
p_3&=&-\frac{b^3}{v_3w_1w_2}\,
\tfrac{\left|\matrix{w_1&w_2\cr u_1^3&u_2^3}\right|}
{  \left|\matrix{1&1\cr u_1&u_2}\right|}\,,
\qquad
q_3=-\frac{v_3w_1w_2}{b^3}\,
\tfrac{\left|\matrix{1&1\cr u_1&u_2}\right|}
{  \left|\matrix{w_1&w_2\cr u_1^2&u_2^2}\right|}\,.
\eea
%
%
\section*{Appendix C}
\renewcommand{\theequation}{C.\arabic{equation}}
\setcounter{equation}{0}
Here we give explicit formulae for the inverse map ${\mathcal S}_3^{-1}$
for the 3-particle periodic Toda lattice (cf. Section 8).
Denote $w_i=\e^{-v_i}$, $i=1,2$, then
\bea
p_1&=&-u_3v_3-u_1-u_2,\qquad \;\e^{q_1}=v_3,
\\
p_2&=&\tfrac{\left|\matrix{u_1w_1&u_2w_2\cr 1&1}\right|}
{  \left|\matrix{w_1&w_2\cr 1&1}\right|}\,,
\qquad \quad
\e^{q_2}=v_3\,
\tfrac{\left|\matrix{w_1&w_2\cr 1&1}\right|}
{  \left|\matrix{1&1\cr u_1&u_2}\right|}\,,
\\
p_3&=&
\tfrac{\left|\matrix{w_1&w_2\cr u_1&u_2}\right|}
{  \left|\matrix{w_1&w_2\cr 1&1}\right|}\,,
\qquad \qquad\quad
\e^{q_3}=-v_3w_1w_2\,
\tfrac{\left|\matrix{1&1\cr u_1&u_2}\right|}
{  \left|\matrix{w_1&w_2\cr 1&1}\right|}\,.
\eea
\pagebreak

\end{document}